\documentclass[reprint,superscriptaddress,amsmath,amssymb,aps,prl,floatfix]{revtex4-1}
\usepackage{graphicx}
\usepackage{bm}
\usepackage{hyperref}
\usepackage{todonotes}
\usepackage{nicefrac}

\definecolor{fj_color}{cmyk}{1, 0.3, 0, 0}
\definecolor{cwh_color}{cmyk}{0, 0.8, 0.8, 0}

\begin{document}
\newcommand{\SRO}{Sr\textsubscript{2}RuO\textsubscript{4}}

\title{Upper Critical Field of \SRO{} under In-Plane Uniaxial Pressure}

\author{Fabian Jerzembeck}
\email{Fabian.Jerzembeck@cpfs.mpg.de}
\affiliation{Max Planck Institute for Chemical Physics of Solids, N\"{o}thnitzer Str 40, 01187 Dresden, Germany}
\author{Alexander Steppke}
\affiliation{Max Planck Institute for Chemical Physics of Solids, N\"{o}thnitzer Str 40, 01187 Dresden, Germany}
\author{Andrej Pustogow}
\affiliation{Department of Physics and Astronomy, University of California Los Angeles, Los Angeles, CA, USA}
\affiliation{Institute of Solid State Physics, TU Wien, 1040 Vienna, Austria}
\author{Yongkang Luo}
\affiliation{Department of Physics and Astronomy, University of California Los Angeles, Los Angeles, CA, USA}
\author{Aaron Chronister}
\affiliation{Department of Physics and Astronomy, University of California Los Angeles, Los Angeles, CA, USA}
\author{Dmitry A. Sokolov}
\affiliation{Max Planck Institute for Chemical Physics of Solids, N\"{o}thnitzer Str 40, 01187 Dresden, Germany}
\author{Naoki Kikugawa}
\affiliation{National Institute for Materials Science, Tsukuba 305-0003, Japan}
\author{You-Sheng Li}
\affiliation{Max Planck Institute for Chemical Physics of Solids, N\"{o}thnitzer Str 40, 01187 Dresden, Germany}
\author{Michael Nicklas}
\affiliation{Max Planck Institute for Chemical Physics of Solids, N\"{o}thnitzer Str 40, 01187 Dresden, Germany}
\author{Stuart E. Brown}
\affiliation{Department of Physics and Astronomy, University of California Los Angeles, Los Angeles, CA, USA}
\author{Andrew P. Mackenzie}
\affiliation{Max Planck Institute for Chemical Physics of Solids, N\"{o}thnitzer Str 40, 01187 Dresden, Germany}
\affiliation{Scottish Universities Physics Alliance, School of Physics and Astronomy, University of St. Andrews, St. Andrews KY16 9SS, United Kingdom}
\author{Clifford W. Hicks}
\affiliation{Max Planck Institute for Chemical Physics of Solids, N\"{o}thnitzer Str 40, 01187 Dresden, Germany}
\affiliation{School of Physics and Astronomy, University of Birmingham, Birmingham B15 2TT, United Kingdom}

\date{\today}

\begin{abstract}

In-plane uniaxial pressure has been shown to strongly tune the superconducting state of \SRO{} by approaching a Lifshitz transition and associated Van Hove singularity (VHS) in the density of states. 
At the VHS, $T_\text{c}$ and the in- and out-of-plane upper critical fields are all strongly enhanced, and the latter has changed its curvature as a function of temperature from convex to concave. 
However, due to strain inhomogeneity it has not been possible so far to determine how the upper critical fields change with strain. 
Here, we show the strain dependence of both upper critical fields, which was achieved due to an improved sample preparation. 
We find that the in-plane upper critical field is mostly linear in $T_\text{c}$. 
On the other hand, the out-of-plane upper critical field varies with a higher power in $T_\text{c}$, and peaks strongly at the VHS. 
The strong increase in magnitude and the change in form of $H_\mathrm{c2||c}$ occur very close to the Van Hove strain, and points to a strong enhancement of both the density of states and the gap magnitude at the Lifshitz transition.
\end{abstract}
\maketitle

Uniaxial pressure affects the superconducting state of \SRO{} strongly \cite{Hicks14_Science, Steppke17_Science, Jerzembeck22_NatComm}.
Applied along a $\langle 100 \rangle$-axis, uniaxial pressure tunes one of the Fermi sheets (the $\gamma$ sheet) through a Lifshitz transition and associated Van Hove singularity in the density of states, at an applied strain of $\varepsilon_\text{VHS} = -0.0044$ (Fig~\ref{fig:setup}(a)) \cite{Sunko19_npj, Barber19_PRB}. 
On the approach to the VHS, $T_\text{c}$ increases quadratically at low strains \cite{Watson18_PRB} and peaks at the VHS, where $T_\text{c}$ is enhanced by a factor of 2.3 \cite{Hicks14_Science, Steppke17_Science, Barber18_PRL}. 
At the VHS, both the in-plane upper critical field, $H_\mathrm{c2 || ab}$, and the out-of-plane upper critical field, $H_\mathrm{c2||c}$, are strongly enhanced: the Pauli limited in-plane upper critical field by a factor of 3 and the orbitally-limited out-of-plane upper critical fields by a factor of 20 \cite{Steppke17_Science}. 
The strengthening of the superconducting state at the Lifshitz transition is associated with a strong enhancement of the total electronic density of states (the Fermi velocity at the Van Hove point goes to zero), resulting in a strong peak in $T_\text{c}$. 
However, a strong enhancement of the orbitally limited upper critical field, $H_\text{c2} \propto (T_\text{c}/v_\text{F})^2$, is only expected when a small Fermi velocity, $v_\text{F}$, coincides with regions of non-zero superconducting gap. 
Hence, the strong enhancement of $H_\mathrm{c2||c}$ was taken as evidence that the superconducting gap is non-zero at the Van Hove point. 
This finding was recently supported by measurements of the heat capacity and the elastocaloric effect under uniaxial pressure \cite{Li21_PNAS, Li22_Science}.

A further intriguing observation is that at the Van Hove singularity, the curvature of the out-of-plane upper critical field changes from a convex (Werthamer-Helfand-Hohenberg-like) form \cite{Werthamer66_PR}, as seen for most superconductors, to a concave form \cite{Steppke17_Science, Li21_PNAS}.  
Among the rare cases of superconductors which show a concave upper critical field are the multi-band superconductors MgB$_2$ \cite{Lyard02_PRB} and Ba(Fe$_{1-x}$Co$_{x}$)$_2$As$_2$ \cite{Ni08_PRB, Kano09_JPSJ}.
A concave form of $H_\text{c2}(T)$ is discussed as an indication of large gap non-uniformity in a single or multiple bands \cite{Shulga98_PRL}.

Here, we aim to get a sense of how the strong changes in the upper critical fields evolve as a function of strain as the Lifshitz transition is approached. 
This technically challenging experiment was accomplished by improving our experimental setup, and hence reducing the effects of strain inhomogeneity noticeably. 
We find that the curvature of $H_\mathrm{c2||c}(T)$ changes only very close to the Van Hove strain. 
Furthermore, we find that the out-of-plane upper critical field varies over a large range in strain with a quadratic power in $T_\text{c}$, as expected for an orbitally-limited critical field. 
In contrast, the in-plane upper critical field $H_{c2||ab}$ is linear in $T_\text{c}$, as expected for a Pauli limited critical field. 
Close to the Van Hove singularity, both $H_\mathrm{c2||ab}$ and $H_\mathrm{c2||c}$ deviate from these power laws and exhibit an overall strong enhancement. 
These strong enhancements indicate that at the Van Hove point not only the density of states but also the superconducting gap magnitude is strongly enhanced.

\section{Experimental results}

\begin{figure}[ptb]
\includegraphics{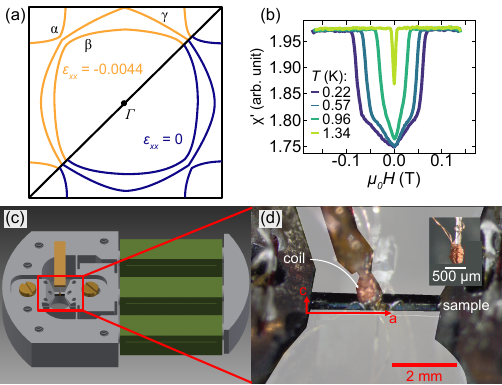}
\caption{(\textbf{a}) Cross-section at $k_z = 0$ of calculated two-dimensional Fermi surfaces at zero strain $\varepsilon_{xx} = 0$ and the Van Hove strain $\varepsilon_{xx} = -0.0044$. (\textbf{b}) Field sweeps at constant temperature of 0.22, 0.57, 0.96 and 1.34~K of a piece of the same crystal as sample 1 and 2. The sharpness of the superconducting transition at low temperatures points to a high-quality sample. The asymmetry of $\chi(H)$ is associated with remanent field effects due to trapped flux in the superconducting magnet. (\textbf{c}) Illustration of the uniaxial stress cell used in this work, and (\textbf{d}) Photograph of sample 2 mounted in this cell.  The inset shows the pair of concentric coils, with one wound directly on top of the other, in more detail.}
\label{fig:setup}
\end{figure}

In previous work on uniaxial pressure tuning, it was observed that, sharp superconducting transitions were only seen near zero pressure and near the Van Hove pressure, where $T_\text{c}$ depends weakly on pressure.
In contrast, at intermediate pressures, where the pressure dependence of $T_\text{c}$ is stronger, the transitions were considerably broadened due to strain inhomogeneity in the samples (The width of the transition is proportional to $dT_\text{c}/d\varepsilon$ and hence the transition broadens away from zero and the Van Hove strain.).  
In order to perform a meaningful study of the critical fields in this intermediate strain region, we took several steps to reduce the effect of strain inhomogeneity.
First of all, we screened multiple samples from different growths by ac susceptibility to find suitable samples (large $T_\text{c}$ and a narrow transition, indicating low internal strain inhomogeneity).
All the samples we investigated were grown by a floating-zone technique \cite{Bobowski19_CondMat} and showed a $T_\text{c}$ close to the clean-limit value \cite{Mackenzie98_PRL}.
Figure \ref{fig:setup}(b) shows ac susceptibility data of a piece of the same rod from which samples 1 and 2 were taken.
The sharpness of the superconducting transitions in magnetic field, down to low temperatures, points to high quality of the crystal, with no apparent effect of ruthenium inclusions \cite{Maeno98_PRL}.
In a second step, we shrank the size of the ac susceptibility coils so that only the most-homogeneously strained region in the center of the sample was probed.
We used a pair of concentric coils with a diameter of $\approx~330~\mu$m, which was placed on top of the sample (Fig. \ref{fig:setup}(c)) with the ac field along the $c$-axis.
Finally, we used samples with high length-to-width and length-to-thickness ratios, reducing sensitivity to the end regions where the applied strain is inhomogeneous.
The bars were mounted in a piezoelectric-based uniaxial pressure cell, as described elsewhere \cite{Hicks14_RSI}.

\begin{figure}[ptb]
\includegraphics{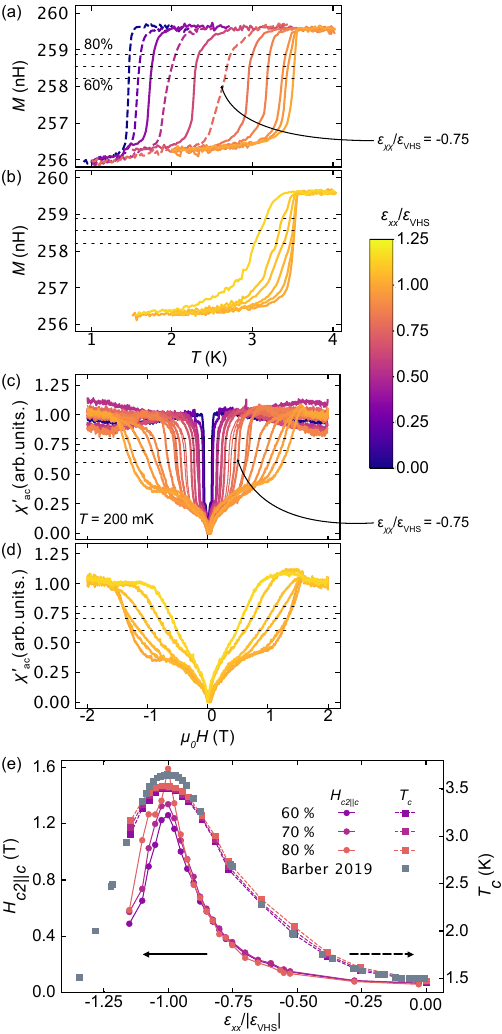}
\caption{Mutual inductance against temperature at a series of compressive strains before (\textbf{a}) and after (\textbf{b}) the Van Hove strain, $\varepsilon_\text{VHS}$, for sample $2$. The dashed curves were measured for increasing $|\varepsilon_{xx}|$ and the full curves for decreasing $|\varepsilon_{xx}|$. (\textbf{c},\textbf{d}) Diamagnetic response against applied field along the crystalline $b$ axis at $200$~mK at a series of compressive strains before and after $\varepsilon_\text{VHS}$. The data are normalized after background correction, as described in detail in the Appendix. The black dashed lines are $60~\%$, $70~\%$ and $80~\%$ thresholds. (\textbf{e}) $T_\text{c}$ (squares) and $H_\mathrm{c2}$ at $200$~mK (dots) against $\varepsilon_{xx}/|\varepsilon_{\mathrm{VHS}}|$. The colors represent the criteria defined in panels (\textbf{a--d}). The grey squares are taken from \cite{Barber19_PRB}.}
\label{fig:exp_data}
\end{figure}

We measured $T_\text{c}$ and $H_\mathrm{c2||c}$ of two samples at a series of compressive strains $\varepsilon_{xx}$.
Figures \ref{fig:exp_data}(a) and (b) show the mutual inductance $M$ between the two coils plotted against temperature of sample $2$ for $\varepsilon < \varepsilon_\text{VHS}$ and $\varepsilon > \varepsilon_\text{VHS}$, respectively.
Sample $1$, which gave similar results as sample $2$, broke at the Van Hove strain, and results are shown in the Appendix.
In contrast, sample $2$ could be compressed to well beyond the Van Hove singularity, and exhibited a better stress homogeneity.
At zero strain, sample $2$ exhibits a sharp transition into the superconducting state at around $1.45$~K, pointing to the high quality of the sample, as already seen in a previous heat capacity measurement on the same sample \cite{Li21_PNAS}.
With increasing uniaxial pressure, $T_\text{c}$ shifts to larger temperature and peaks at the Van Hove strain before falling steeply.
The superconducting transition broadens slightly, but remains narrower than in previous experiments \cite{Steppke17_Science, Barber18_PRL}.
In order to verify that the sample and the epoxy remained within their elastic limits, $T_\text{c}$ was determined both before and after the sample was taken to maximum pressure, and no substantial difference was observed.
However, as can be seen in Figure~\ref{fig:exp_data}(a), the superconducting transition for increasing (dashed lines) and decreasing $|\varepsilon_{xx}|$ (full lines) showed differences: the former is broader than the latter.
We attribute this difference to minor fracture of the epoxy that reduced sample bending when stress was applied.

Figures \ref{fig:exp_data}(c) and (d) show the diamagnetic response against field applied along the $c$ axis at $200$~mK.
A fourth-order polynomial background is subtracted from the data; details are given in the Appendix.
We will focus again on sample $2$ and show data from sample $1$, which gave similar results, in the Appendix.
For small fields, a weak increase of the diamagnetic response is visible, which is associated with vortex motion. At higher fields a sharp superconducting transition occurs at around 67~mT for sample 2 at zero strain.
With increasing compressive strain, the superconducting transition shifts to larger fields.
The transition broadens, but remains much narrower than in previous measurements, allowing us to determine the strain dependence of $H_\text{c2}$ for the first time. 
At the Van Hove strain, $H_\mathrm{c2||c}$ is enhanced by a factor of $\approx 19$, in good agreement with previous results \cite{Steppke17_Science}. 
Beyond $\varepsilon_{\mathrm{VHS}}$, $H_\mathrm{c2||c}$ falls steeply.

We now compare the strain dependences of $H_\text{c2}$ and $T_\text{c}$.
Figure~\ref{fig:exp_data}(e) shows $T_\text{c}$ (squares) and $H_\mathrm{c2||c}$ (dots) of sample $2$ against $\varepsilon_{xx}/|\varepsilon_{\mathrm{VHS}}|$.
$H_\text{c2}$ is best identified by the onset of the superconducting transition. 
However, due to transition broadening, a threshold criterion is more practical.
Hence, the colors represent the $60~\%$, $70~\%$ and $80~\%$ levels, which are marked by dashed lines in Panels~\ref{fig:exp_data}(a)-(d).
Both $T_\text{c}$ and $H_\mathrm{c2||c}$ peak, within the resolution of the experiment, at the same Van Hove strain.
$T_\text{c}$ increases approximately quadratically at low strains and shows a broad peak around the Van Hove strain.
The width of the peak in $T_\text{c}$ is similar to that observed previously \cite{Barber19_PRB}.
In contrast, the peak in $H_\text{c2}$ is very sharp.
The much narrower peak of $H_\mathrm{c2||c}$ compared to $T_\text{c}$ might be a temperature effect: $H_\mathrm{c2||c}$ of sample $2$ was measured at $200$~mK, whereas $T_\text{c}$ was measured between $1.5$~K and $3.5$~K.
In comparison, the peak of $H_\mathrm{c2||c}$ of sample $1$, shown in the Appendix and measured at $900$~mK, is broader than the the peak of sample $2$ but still narrower that the peak of $T_\text{c}$.
The width of the peak in $H_\mathrm{c2||c}$ sets an upper limit on strain inhomogeneity, which means that the observed peak width for $T_\text{c}$ is intrinsic.

\begin{figure}[ptb]
\includegraphics{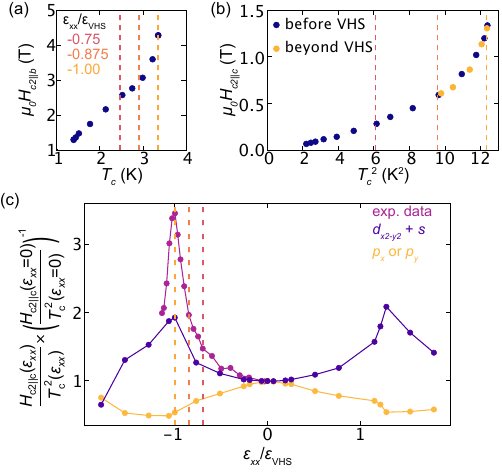}
\caption{(\textbf{a}) $H_\mathrm{c2||b}$ at 20~mK against $T_\text{c}$ up to the Van Hove strain. (\textbf{b}) $H_\mathrm{c2||c}$ against $T_\text{c}^2$ for strains before (blue dots) and beyond (yellow dots) the Van Hove strain. (\textbf{c}) $H_\mathrm{c2||c}/T_\text{c}^2$ against $\varepsilon_{xx}/\varepsilon_\mathrm{VHS}$. For comparison, results from weak-coupling calculations for an even- and an odd-parity order parameter, taken from Ref.~\cite{Steppke17_Science}, are also shown. The colored lines represent the $\varepsilon_{xx}/\varepsilon_\mathrm{VHS} = 0.75, 0.875$ and $1$ values.}
\label{fig:main_results}
\end{figure}

Next we turn to the strain dependence of the in-plane upper critical field, $H_\mathrm{c2||b}$, which was measured by ac susceptibility at $20$~mK.
Experimental details can be found in Ref.~\cite{Pustogow19_Nature}.
Figure~\ref{fig:main_results}(a) shows $H_\mathrm{c2||b}$ against $T_\text{c}$.
The in-plane upper critical field is approximately proportional to $T_\text{c}$ and deviates from proportionality only very close to the Lifshitz transition.
The linear dependence of $H_\mathrm{c2||b}$ on $T_\text{c}$, and therefore on the $k$-averaged value of $\Delta(k)$, is expected for a Pauli limited critical field \cite{Clogston62_PRL}.
Pauli limiting is associated with spin-singlet superconductivity, so this observation is consistent with an even-parity state \cite{Pustogow19_Nature, Ishida20_JPSJ, Petsch20_PRL, Chronister21_PNAS}, and results in a first-order transition, which has been observed both at zero strain \cite{Yonezawa13_PRL, Kinjo22_Science} and at the Van Hove strain \cite{Steppke17_Science}.

In order to further understand how the superconducting state evolves as the Lifshitz transition is approached, we plot the orbitally limited $H_\mathrm{c2||c}$ against $T_\text{c}^2$ in Fig.~\ref{fig:main_results}(b). 
If, hypothetically, the gap of a superconductor is scaled without modification of its $k$-space structure, $H_\text{c2} \propto T_c^2$ is expected, and the constant of proportionality is proportional to the density of states squared \cite{Steppke17_Science}.
Figure \ref{fig:main_results}(b) shows the out-of-plane upper critical field against $T_\text{c}^2$ of sample $2$ for compressive strains before (blue dots) and after (yellow dots) the Van Hove singularity.
For small values of $T_\text{c}$, $H_\mathrm{c2||c}$ increases only slightly faster than $T_\text{c}^2$.
But close to the Van Hove strain, $H_\mathrm{c2||c}$ deviates further from the $T_\text{c}^2$ dependence and exhibits overall a super-quadratic dependence in $T_\text{c}$, resulting in a sharp rise of $H_\mathrm{c2||c}$ over an already strong enhancement.
This behaviour indicates that, relative to the unstrained material, gap weight shifts to sections of Fermi surface where the Fermi velocity is lower and the density of states is higher. 

Finally, by re-plotting the data as $H_\mathrm{c2||c}/T_\text{c}^2$ against strain in Figure \ref{fig:main_results}(c), we can compare the experimental results to predictions from two-dimensional weak-coupling calculations, taken from Ref.~\cite{Steppke17_Science}, for even- ($d_{x^2-y^2} + s$) and odd-parity ($p_x$ or $p_y$) order parameters.
As noted above, the increase in $H_\mathrm{c2||c}/T_\text{c}^2$ indicates a non-zero gap in the vicinity of the Lifshitz transition, which is in two dimensions only possible for even-parity order (In three dimensions, a finite gap can only occur for odd parity order parameters with horizontal line nodes) \cite{GrgurPalle}.
It is notable that the observed $H_\mathrm{c2||c}/T_\text{c}^2$ peaks close to the VHS much more sharply than in the calculation.
At the Van Hove singularity, $H_\mathrm{c2||c}/T_\text{c}^2$ is enhanced by a factor of $\approx 3.5$, in good agreement with Ref.~\cite{Steppke17_Science}.
The much larger enhancement of $H_\mathrm{c2||c}/T_\text{c}^2$ over the calculations might be explained by strengthened many-body effects as pointed out in Ref.~\cite{Steppke17_Science} and discussed in \cite{Luo19_PRX}.

\begin{figure}[ptb]
\includegraphics{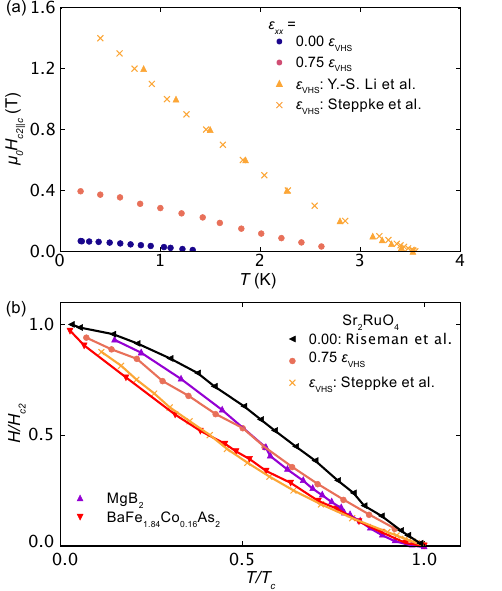}
\caption{(\textbf{a}) $H_\mathrm{c2||c}$ against temperature for three different compressive strains. Dots are measured by ac susceptibility and triangle by heat capacity, on the same sample \cite{Li21_PNAS}. X's are ac susceptibility data taken from \cite{Steppke17_Science}. (\textbf{b}) Comparison of the change in curvature of \SRO{} at different compressive strains with the concave $H_\text{c2}(T)$ curves of MgB$_2$ \cite{Lyard02_PRB} and BaFe$_{1.84}$Co$_{0.16}$As$_2$ \cite{Kano09_JPSJ}. Due to the larger temperature range, the zero strain \SRO{} is taken from \cite{Riseman98_Nature}.}
\label{fig:Hc2(T)}
\end{figure}

Next, we turn to the temperature dependence of the upper critical field at intermediate strains.
In previous studies \cite{Steppke17_Science, Li21_PNAS} it was found that $H_\mathrm{c2||c}(T)$ changes from a convex function of temperature at zero strain to a concave form at the Van Hove strain.
Figure~\ref{fig:Hc2(T)}(a) shows $H_\mathrm{c2||c}(T)$ measured by ac susceptibility and heat capacity of sample $2$ at three compressional strains.
Details about the heat capacity measurements can be found in Ref.~\cite{Li20_RSI, Li21_PNAS}.
The data show that the change of $H_\mathrm{c2||c}(T)$ from a convex function of temperature to a concave function occurs close to the Van Hove strain: At $\varepsilon_{xx}/\varepsilon_{\mathrm{VHS}} = 0.75$, $H_\mathrm{c2}(T)$ is still convex.
In other words, the change from a convex to a concave shape only occurs in a similar range of strain to that over which $H_\mathrm{c2||c}$ deviates strongly from a $T_c^2$ dependence.

In Figure~\ref{fig:Hc2(T)}(b) we plot our $H_\mathrm{c2}(T,\varepsilon)$ under strain data normalised to their respective $T_\text{c}$ and $H_\mathrm{c2}(T=0)$ values, and compare them to the $H_\mathrm{c2}(T)$-curves of the multi-band superconductors MgB$_2$ and overdoped  BaFe$_{1.84}$Co$_{0.16}$As$_2$ \cite{Lyard02_PRB, Kano09_JPSJ}.
Instead of our own zero-strain $H_\mathrm{c2}(T)$ curve, we show the curve from Ref.~\cite{Riseman98_Nature} due to the larger temperature range. 
However, it is noteworthy that over the measured temperature range there is no essential difference between the two zero-strain \SRO{} $H_{c2}(T)$ curves.
At the Van Hove strain, where $H_\mathrm{c2}(T)$ of \SRO{} has changed its curvature from convex to concave, $H_\mathrm{c2}(T)$ matches the curve of BaFe$_{1.84}$Co$_{0.16}$As$_2$ very closely, and is even more concave than that of the textbook two-gap superconductor MgB$_2$.

The concave temperature dependence of $H_\mathrm{c2||c}(T)$ raises the question of whether we are actually measuring the upper critical field or rather the so-called irreversibility line, which is associated with the melting of the flux lattice \cite{Gammel88_PRL, Osofsky93_PRL, Fuchs01_SSC} at a first-order phase transition \cite{Schilling96_Nature}.
Flux lattice melting usually occurs in quasi-two-dimensional superconductors with a short coherence lengths and high superconducting transition temperatures.  
Even at the Van Hove strain, the coherence length of \SRO{} is quite long, approximately 200 angstroms.  
Also, the transitions in susceptibility and heat capacity are sharp and the critical fields deduced from the two measurements are in good agreement (Fig.~\ref{fig:Hc2(T)}b), and there is no experimental evidence of either a first order transition or of substantial diamagnetic fluctuations above $T_\text{c}(H)$. 
We therefore conclude that our observed $H_\mathrm{c2}(T)$ curve is that of the thermodynamic order parameter, and not a consequence of flux lattice melting.

A concave $H_\text{c2}(T)$ has also been discussed in the context of quantum critical points \cite{Kotliar96_PRL}, order parameter mixing \cite{Koyama96_PhysC} and the proximity to a VHS \cite{Dias96_SSC}. However, the latter predicts a relation $H_{c2} \propto T_\text{c}^{\sqrt{2}}$, which is not observed in our experiments.
In summary, the microscopic details of the change in curvature of $H_\mathrm{c2}(T)$ of \SRO{} are not well understood. 
We hope that this finding will motivate future work to understand the concave nature of $H_{c2}(T)$.

\section{Discussion}

We have shown that by $\langle 100 \rangle$ uniaxial pressure tuning \SRO{} to a Lifshitz transition and associated Van Hove singularity, the orbitally-limited $H_\mathrm{c2||c}$ exhibits a sharp rise over an already strong enhancement, pointing to a large gap coinciding with a small Fermi velocity at $\varepsilon_\text{VHS}$.
In a similar range of strains as the strong enhancement, $H_\mathrm{c2||c}$ also changes its form from convex to concave.
These sudden changes indicate that the out-of-plane upper critical field is highly sensitive to something occurring around the Van Hove strain.
On the other hand, the in-plane upper critical field is Pauli limited and exhibits a linear dependence in $T_\text{c}$. 
In the simplest single-band situation ($g = 2$), the Pauli limited field is given by $H_{\mathrm{P}} = \Delta(0)/(\sqrt{2S}\mu_{\mathrm{B}})$, where $S = (1-VN(E_{\text{F}}))^{-1}$ is the renormalization due to the Stoner factor \cite{Clogston62_PRL}. 
For materials with $T_\text{c} = 1.5$ K and an isotropic gap, a Pauli-limited critical field of $2.76$~T is expected. 
This is about twice the value observed for unstressed \SRO{}, suggesting that the Stoner factor is substantial.
On the approach to the Lifshitz transition, Knight shift measurements \cite{Luo19_PRX, Chronister2022_npjQM} and DFT calculations found a continuously and strongly increasing density of states \cite{Steppke17_Science}, which would naturally result in a gradually increasing Stoner factor.
Indeed, at the Van Hove strain the Stoner factor is enhanced by $\approx 30~\%$ over the zero strain value \cite{Luo19_PRX}.
As a consequence the Pauli limited field should be gradually suppressed, resulting in a  sub-linear dependence of the in-plane upper critical field on strain. 
Since this sub-linear behavior is not observed, the superconducting gap must increase faster than linearly in $T_\text{c}$ to compensate the suppression due to the Stoner factor, resulting in an overall quasi-linear dependence of the Pauli limited field in $T_\text{c}$.
A strengthening of the superconducting gap might also explain the sudden upturn of $H_\mathrm{c2||b}$, which occurs at a similar strain as the changes in $H_\mathrm{c2||c}$. However, since this upturn is not criterion-independent (Fig.~\ref{fig:NMR}), more studies are needed to clarify this hypothesis. 
Finally, the strong enhancement of both the density of states and the gap magnitude close to the Van Hove singularity could also explain the large difference between the experimental values and the weak-coupling calculation of $H_\mathrm{c2||c}/T_\text{c}^2$ close to the Van Hove strain, shown in Figure~\ref{fig:main_results}(c).

In summary, we have determined the strain dependence of the upper critical fields of \SRO{} for uniaxial pressures along the $a$-axis between zero strain and the Van Hove strain.
This was achieved by an improved sample preparation process and a size reduction of the susceptometer, which overcame some of the challenges of strain inhomogeneity.
We find that the in-plane upper critical field exhibits a linear dependence in $T_\text{c}$, expected for a Pauli limited field $H_{\text{P}}$.
On the other hand, the out-of-plane critical field peaks much more sharply than $T_\text{c}$ on approaching the Van Hove strain, which points to a large superconducting gap coinciding with a small Fermi velocity.
At a similar strain to the sudden rise in $H_\mathrm{c2}$, the temperature dependence of the out-of-plane upper critical field changes from a convex to a concave form.
The dramatic changes in the electronic structure and the superconducting properties occurs close to the Van Hove strain, which implies a large sensitivity of the upper critical fields to the Lifshitz transition.  
Our findings motivate careful study in this range of strain, studying the critical fields at finely spaced strain values.

\section{Appendix}

\begin{figure}[ptb]
\includegraphics{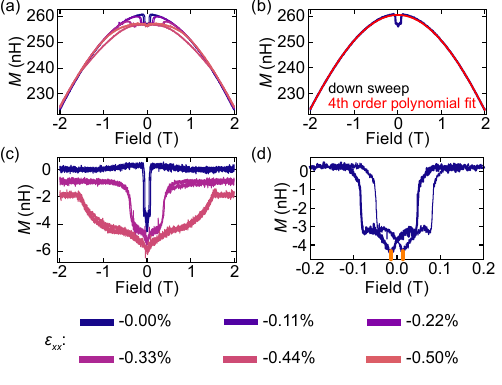}
\caption{(\textbf{a}) Mutual inductance against applied magnetic field for a series of compressive strains for sample $2$. (\textbf{b}) $4$th order polynomial fit to an individual sweep over the fitting range $|\mu_0H| > 1.6$~T. (\textbf{c}) Data after subtracting the polynomial fit. (\textbf{d}) A close-up of data in panel (c), showing the magnet hysteresis (orange bars). The strains are normalized, so that the Van Hove strain is in agreement with the literature value, $\epsilon_\text{VHS} = -0.0044$ \cite{Barber19_PRB}.}
\label{fig:field_corr}
\end{figure}

\textit{Background subtraction.}
In field sweeps, there was a strongly-varying background signal, as shown in Fig.~\ref{fig:field_corr}(a).
This background signal was frequency-dependent but almost independent of temperature, and present above and below $T_\text{c}$.
It differed in magnitude and sign for both samples, showing that it is an artifact of interaction of the sense coils with the applied field, and not intrinsic to \SRO{}.
Figure~\ref{fig:field_corr}(a) shows raw data from sample $2$ for applied fields $-2~\mathrm{T} < \mu_0H < 2~\mathrm{T}$ at a series of compressive strains. 
In addition to the background signal, a small hysteresis is apparent between field-up and down sweeps, due to flux pinning in the magnet.
Despite this background, the superconducting transition is visible for small strains.
With increasing strain the transition shifts to larger fields until it is barely visible for the high strain data.
In order to subtract the background signal, a fourth-order polynomial was fitted to data at $|\mu_0H| > 1.6$~T, independently at each strain and for the increasing- and decreasing-field data, as shown in Figure~\ref{fig:field_corr}(b).
Figure~\ref{fig:field_corr}(c) shows data at a series of strains after the background subtraction.
A variation of the normal-state level on the order of $~1$~nH indicates that the background subtraction is not perfect.
After subtracting the background, the hysteresis was corrected by locating the minima in the diamagnetic signal (which is the true zero field and is indicated by orange bars in Figure~\ref{fig:field_corr}(d)), and subtracting the field associated with the minimum.
In a final step, the curves were all normalized independently.

\begin{figure}[ptb]
\includegraphics{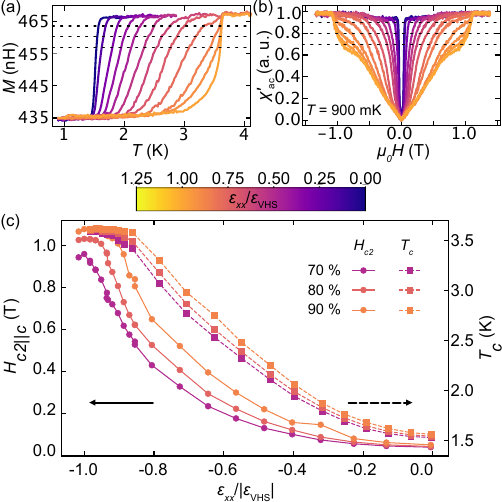}
\caption{Diamagnetic response against temperature (\textbf{a}) and applied field (\textbf{b}) at a series of compressive strains up to the Van Hove strain for sample $1$. For the field sweeps, the background was subtracted as described in detail in the Appendix. The black dashed lines are a $70~\%$, $80~\%$ and $90~\%$ threshold. (\textbf{c}) $T_c$ (squares) and $H_{c2}$ at $900$~mK (dots) against $\varepsilon_{xx}/|\varepsilon_{\mathrm{VHS}}|$. The colors represent the criteria defined in panels (\textbf{a--b}).}
\label{fig:sample-1}
\end{figure}

\textit{Data from sample 1.}
As noted in the main text, sample 1 had lower strain homogeneity than sample 2, and broke at the Van Hove strain.
Figure~\ref{fig:sample-1} shows ac susceptibility data as a function of temperature (a) and applied $c$-axis field (b) at a series of compressive strains up to the Van Hove strain.
The data taken in field sweeps were analyzed with the same procedure as for sample $2$.
For this sample a different cryostat was used, which could apply a maximum field of only $1.5$~T. 
Since the zero-temperature upper critical field at the Van Hove singularity is $\approx 1.5$~T, the strain dependence of $H_\mathrm{c2||c}$ was determined at $900$~mK.
With increasing strain, $H_\mathrm{c2||c}$ shifts to large fields and the transition broadens due to strain inhomogeneity.
Close to the Van Hove strain the superconducting transition sharpens due to the small $dT_\text{c}/d\varepsilon$ or $dH_\mathrm{c2}/d\varepsilon$.
For sample $1$, the transition sharpens noticeably in field sweeps close to the normal state level, pointing to overall larger strain inhomogeneity than sample 2.
Hence, we chose a $70~\%$, $80~\%$ and $90~\%$ criteria to determine $H_{c2}$ and $T_\text{c}$.
Figure~\ref{fig:sample-1}(c) shows $T_\text{c}$ (squares) and $H_\mathrm{c2||c}$ at $900$~mK (dots) against strain, normalized by the Van Hove strain.

\begin{figure}[ptb]
\includegraphics{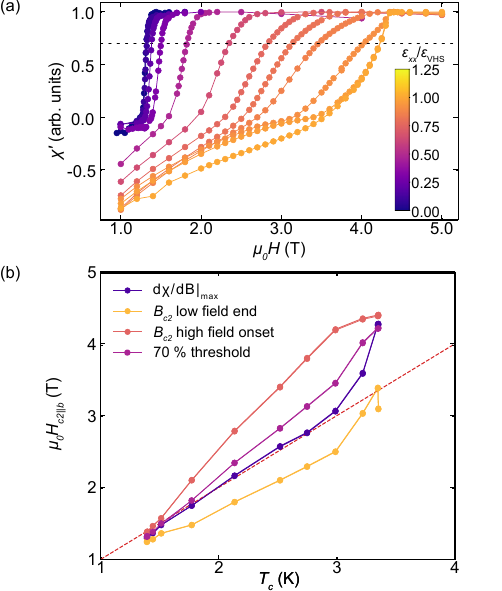}
\caption{(\textbf{a}) Magnetic susceptibility against magnetic field, applied parallel to the crystalline $b$ axis, at a series of compressive strains. The data was taken from Ref.~\cite{Pustogow19_Nature}. The curves were individually normalized. The dashed line is a 70~\% criterion. (\textbf{b}) $H_\mathrm{c2||b}$ for different criteria at 20~mK against $T_\text{c}$ up to the Van Hove strain. The maximum in the slope of the transition (dark blue) was taken as a criterion in Fig.~\ref{fig:main_results}. In comparison, an onset criterion for the transition at low fields (yellow) and high fields (red) and the 70~\% criterion (purple) from panel (\textbf{a}) are shown.}
\label{fig:NMR}
\end{figure}

\textit{Additional data.}
Finally, we present in Figure~\ref{fig:NMR}(a) the magnetic susceptibility against field applied along the in-plane $b$ axis at 20~mK.
For better visibility, the curves were individual normalized.
The data was already presented in the Extended Data of Ref.~\cite{Pustogow19_Nature}.
Figure~\ref{fig:NMR}(b) shows $H_\mathrm{c2||b}$ against $T_\text{c}$ for different criteria for the upper critical field.
The in-plane upper critical field is determined by the maximum slope in the transition (purple) and by a low-end (yellow) and onset (red) criteria of the transition, as defined in the Extended Data in Ref.~\cite{Pustogow19_Nature}, and additionally by a 70~\% threshold criterion, defined by the dashed line in Panel~(\textbf{b}).
All criteria find a linear dependence of $H_\mathrm{c2||b}$ in $T_\text{c}$ for a large range of strains, as expected for a Pauli limited field.
Close to the Lifshitz transition, $H_\mathrm{c2||b}$ exhibits a non-linear behavior in $T_\text{c}$ for all criteria, but the precise form of this non-linearity is criterion-dependent, and so has not been firmly established.
At the Van Hove singularity, the onset, the threshold and the $dM/dB|_\text{max}$ criteria approach the same value with $H_\mathrm{c2||b}/T_\text{c} > 1$, as found in a numerical study for even-parity order parameters \cite{Yu20_PRB} and in agreement with previous results \cite{Steppke17_Science}.

\textit{Acknowledgements.} We thank Thomas Scaffidi, Mark E. Barber, Grgur Palle and Daniel Agterberg for helpful discussions.
F.J., A.P.M., and C.W.H. acknowledge the financial support of the Deutsche Forschungsgemeinschaft (DFG, German Research Foundation) - TRR 288 - 422213477 (project A10). 
A.P. acknowledges support by the Alexander von Humboldt Foundation through a Feodor Lynen Fellowship during his work at UCLA.
NK is supported by a KAKENHI Grants-in-Aids for Scientific Research (Grant Nos. 18K04715, 21H01033, and 22K19093), and Core-to-Core Program (No. JPJSCCA20170002) from the Japan Society for the Promotion of Science (JSPS) and by a JST-Mirai Program (Grant No. JPMJMI18A3).
A.C. is grateful for support from the Julian Schwinger Foundation for Physics Research. A.P. acknowledges support by the Alexander von Humboldt Foundation through the Feodor Lynen Fellowship. The work at University of California, Los Angeles, was supported by NSF Grant 2004553.

\textit{Author contributions} F.J., C.W.H., S.E.B, and A.P.M. designed the research project; F.J., A.S., A.C., Y.L, A.P. and Y.-S.L. performed the measurements; N.K. and D.A.S. grew the samples; F.J. and C.W.H. analyzed data; and F.J. and C.W.H. wrote the paper with contributions from the other authors.

\section{Data Availability}

Raw data are available on website yet to be determined.

\bibliography{bibliography.bib}

\end{document}